\documentclass[runningheads,fleqn]{svmult}

\usepackage{makeidx}   % allows index generation
\usepackage{graphicx}  % standard LaTeX graphics tool
\usepackage{subeqnar}  % subnumbers individual equations
\usepackage{multicol}  % used for the two-column index
\usepackage{physproc}  % flushleft layout of diverse elements,etc.
\makeindex             % used for the subject index
\begin{document}
\title*{Diagrammatic Monte Carlo}
\toctitle{Diagrammatic Monte Carlo}
% allows explicit linebreak for the table of content
%
%
\titlerunning{Diagrammatic Monte Carlo}
% allows abbreviation of title, if the full title is too long
% to fit in the running head
%
\author{Kris Van Houcke  \inst{1,2}
\and Evgeny Kozik \inst{1,3} \and N. Prokof'ev \inst{1,3,4} \and
B. Svistunov \inst{1,4}}
\authorrunning{Kris Van Houcke et al.}
% if there are more than two authors,
% please abbreviate author list for running head
%
%
\institute{Department of Physics, University of Massachusetts,
Amherst, MA 01003, USA \and Universiteit Gent, Vakgroep Subatomaire
en Stralingsfysica,\\ Proeftuinstraat 86, B-9000 Gent, Belgium
\and Institut f\"{u}r Theoretische Physik, ETH Z\"urich, CH-8093 Z\"urich,
Switzerland \and Russian Research Center ``Kurchatov Institute",
123182 Moscow, Russia}
\maketitle              % typesets the title of the contribution

\begin{abstract}
Diagrammatic Monte Carlo (DiagMC) is a numeric technique that allows one
to calculate quantities specified in terms of diagrammatic
expansions, the latter being a standard tool of many-body quantum
statistics.  The sign problem that is typically fatal to Monte
Carlo approaches, appears to be manageable with DiagMC. Starting
with a general introduction to the principles of DiagMC, we present
a detailed description of the DiagMC scheme for interacting fermions
(Hubbard model), as well as the first illustrative results for the
equations of state.
\end{abstract}

\section{Introduction. General Principles}

Diagrammatic expansion (Feynman diagrams) is a powerful generic
tool of quantum statistics \cite{Fetter_Walecka}. Mathematically,
diagrammatic expansion---for some relevant quantity, $Q$, usually
a Green's function---is a series of integrals with an ever
increasing number of integration variables,
\begin{equation}
Q(y) \, = \, \sum_{m=0}^{\infty} \sum_{\xi_m} \int  \mathcal{D}(\xi_m, y,
x_1, \ldots , x_m)\, dx_1 \cdots dx_m  \; . \label{main}
\end{equation}
Here $y$ is a set of parameters which the quantity $Q$ can depend
on,  $\xi_m$ indexes different terms of the same order $m$ (the
term $m=0$ is understood as a function of $y$ only), and
the $x$'s are the integration variables. Structures similar to
Eq.~(\ref{main}) originate from perturbative expansions of the
quantum-statistical averages, in which case the functions $\mathcal{D}$
can be represented by Feynman diagrams/graphs, with the graph
lines standing for either Green's functions (propagators) $G$ or
interaction potentials $U$, so that the whole diagram encodes
a certain product of $G$'s and $U$'s. It will be essential for our
purposes to approach the integrations in Eq.~(\ref{main}) in the same way as
the diagram order $m$ and its topology when defining the notion
of a diagram, in contrast to analytic treatments where integrations
are included into the definition of $\mathcal{D}$.

Diagrammatic Monte Carlo (DiagMC) \cite{DiagMC} is a technique
that allows one to simulate quantities specified in terms of a
diagrammatic series. In a broad sense, it is a set of simple
generic prescriptions for organizing a systematic-error-free
Metropolis-Rosenbluth-Teller type process that
samples the series (\ref{main}) without explicitly performing
integrations over the internal variables in each particular term.
The rules are as follows.

The function $Q(y)$ is interpreted as a distribution function for
the variable(s) $y$.  The statistical interpretation of
Eq.~(\ref{main}) suggests calculating $Q(y)$ by a Markov-chain
process which samples diagrams stochastically. The value of
$\mathcal{D}$ plays the role of the statistical weight of the
diagram (i.e., the probability with which the diagram is
generated). More precisely, to deal with sign-alternating series,
one writes $\mathcal{D}$ as a product of its absolute value
$|\mathcal{D}|$, which determines the statistical weight of the
diagram, and the sign function $s_\mathcal{D}= \pm 1$, which
contributes to the statistics for $Q$ each time the term
$\mathcal{D}$ is generated. Since contributions of different
$m$-th order diagrams are of the same order of magnitude and their
number grows factorially with $m$, the combined $m$-th order
contribution to $Q$ is the result of nearly complete cancellation
of the diagrams of different sign. This is the notorious sign
problem (see, e.g., \cite{Troyer}). In our case, it causes an
exponential scaling of the computation time with the maximum
diagram order, making it impossible to obtain sensible results at
large $m$. At this point one faces the problem of extrapolating
the answer to the $m \to \infty$ limit. To make matters worse, the
series may be asymptotic/divergent. For this reason, the use of
\textit{resummation techniques} \cite{fermipolaron} is a crucial
ingredient of DiagMC (see Secs.~\ref{sec:tricks},
\ref{sec:results}).

Though DiagMC has to confront the sign problem,
just like other exact techniques (e.g., the dynamical
cluster algorithms \cite{DCA}), it has an important advantage: it
works immediately in the thermodynamic limit, and is {\it not} subject to the exponential
scaling of the computational complexity with the system or cluster volume.
DiagMC also has a potential for vast improvements in efficiency
by using standard tricks of the analytic diagrammatic approach,
developed to reduce the number of diagrams calculated explicitly term-by-term
by self-consistently taking into account chains of repeating parts as,
e.g., in the Dyson equation \cite{Fetter_Walecka}. These ideas are the essence of the
Bold Monte Carlo scheme introduced in Ref.~\cite{bmc} and successfully applied
to the Fermi-polaron problem in Ref.~\cite{fermipolaron}.
In Sec.~\ref{sec:tricks}, we describe the basic steps in this direction
for DiagMC, such as the use of \textit{bold propagators}.

The stochastic sampling of $Q(y)$ consists of a number of
elementary sub-processes (or updates) falling into two qualitatively
different classes: (I) updates which do not change the number of continuous variables
(they change the values of variables in $\mathcal{D}$,
but not the form of the function itself), and (II) updates which change the structure of $\mathcal{D}$.
The processes of class I are rather straightforward, being identical to those of simulating a given continuous distribution
$|\mathcal{D}(\xi_m, y, x_1, \ldots , x_m)|$ of the variables $x_1, \ldots , x_m$.

The crucial part is played by type-II updates, arranged to form
complementary pairs with the detailed balance condition \cite{Metropolis} satisfied
in each pair. Let $\cal A$ transform the diagram ${\mathcal D}(\xi_m, y, x_1, \ldots , x_m)$ into
$\mathcal{D}(\xi_{m+n}, y, x_1, \ldots  , x_{m+n})$,
and, correspondingly, its counterpart $\cal B$ performs the inverse
transformation. For the new variables we introduce a vector
notation: ${\vec x} = \{ x_{m+1}, \ldots , x_{m+n} \}$.
The first step in $\cal A$ consists of selecting some type of
diagram transformation and proposing $\vec x$ which is
generated from a certain probability distribution $W(\vec
x)$. The form of $W(\vec x)$ is arbitrary, but to render the algorithm
most efficient, it is desirable that $W(\vec x)$ is (i) of a simple
analytic form with known normalization, and (ii) close to the actual
statistics of $\vec x$ in the final diagram. The second step consists
of accepting the proposal with probability, $P_{\mbox{\scriptsize add}}(\vec
x)$. In $\cal B$ one simply accepts the proposal for removing the
variables $\vec x$ with the probability $P_{\mbox{\scriptsize rem}}(\vec x)$.
The pair of complementary sub-processes is
balanced if the following equality is fulfilled \cite{DiagMC}:
\begin{equation}
P_{\mbox{\scriptsize add}}(\vec x)  \: = \: \min \{
R(\vec x) / W(\vec x)  , 1\} \;,
\label{P_acc}
\end{equation}
\begin{equation}
P_{\mbox{\scriptsize rem}}(\vec x)  \: = \: \min \{
W(\vec x) / R(\vec x)  , 1 \} \;,
\label{P_rem}
\end{equation}
where
\begin{equation}
R(\vec x) \, = \,( p_{\cal B}/p_{\cal A}) \, |\mathcal{D}(\xi_{m+n}, y, x_1,
\ldots , x_m, {\vec x})/ \mathcal{D}(\xi_m, y, x_1,  \ldots , x_m) \label{R}|
\; ,
\end{equation}
with $p_{\cal A}$ and $p_{\cal B}$ being the probabilities of
addressing the sub-processes $\cal A$ and $\cal B$. [Often,
$p_{\cal A} \neq p_{\cal B}$ is a natural choice.]
The above protocol is a straightforward generalization of the standard
approach \cite{Metropolis,Hastings} to the sampling of functions
with a \textit{variable} number of continuous arguments.

Generally speaking, the set of DiagMC updates is specific for
a given type of the diagrammatic expansion, being sensitive to
both the topology and the representation
(say, momentum or coordinate) of the diagrams. The updating
strategy can be rendered more/less sophisticated, depending on
what is being optimized: the simplicity or the efficiency.
Examples of particular DiagMC schemes can be found in
Refs.~\cite{DiagMC,fermipolaron,Mishchenko,Burovski_01,Burovski_06}.

In this work, we consider an interacting many-body Fermi system and work
with diagrams in the imaginary-time--momentum representation.
We employ a sophisticated (at first glance) updating scheme
based on the worm algorithm idea \cite{worm} when
updating flexibility and efficiency are achieved by extending
the configurational space (in our case, the space of allowed diagrams).
As far as we know, the algorithm being presented is the
first application of the DiagMC method to connected many-body
Feynman diagrams.

Perturbative expansions in the interaction are often divergent.
Dyson gave a simple physical argument leading to a
\textit{sufficient} condition for an expansion series to diverge:
if changing the sign (or phase) of some parameter $\gamma$
implies an abrupt change of the physical state
(e.g., the change of symmetry, or an instability),
then $\gamma=0$ is the point of non-analyticity and the expansion in
powers of $\gamma$ is \textit{a priori} divergent \cite{Dyson}.
In particular, this means the diagrammatic
expansion with respect to the interaction in continuous space is
divergent for both bosons and fermions, since both
systems collapse at negative (on average) interaction. Obviously,
the opposite is not necessarily true---that is the absence of
singularity in the aforementioned sense does not guarantee
analyticity (and thus convergence at small enough $\gamma$)---but
a different reason for a series to be non-analytic (and thus
divergent at arbitrarily small  $\gamma$) should be rather exotic.
Therefore,  we shall rely on the physical picture to
predict the analytic properties of the series at $\gamma\to 0$.

However, the real power of the analytic diagrammatic technique lies in the
possibility of reducing infinite sums of diagrams
(no matter convergent or divergent) to simple integral equations, e.g., the Dyson
equation. In addition to that, for divergent asymptotic series one can
employ generalized resummation schemes to obtain the answer outside
of the series convergence radius. Finally, when the resummation trick
fails, the series can be analyzed using approximate (biased) methods,
which assume a certain function behind the series.
An example of the successful application of the resummation method
to a divergent sign alternating series for polarons can be found in Ref.~\cite{fermipolaron}.
In this work we demonstrate that the DiagMC method is a viable approach
to study interacting many-body systems. Even moderate success in this direction
is extremely important since other Monte Carlo approaches face a
severe sign problem before the fully controlled extrapolation to
the thermodynamic limit can be done.

\section{Model and Diagrams }

In what follows we discuss the Fermi-Hubbard model,
\begin{equation}
H\, =\, \sum_{\mathbf{k}, \sigma}\, (\varepsilon_{\mathbf{k}}-
\mu_\sigma) \, a^{\dagger}_{\sigma \mathbf{k}}a^{\phantom{\dagger}}_{\sigma
\mathbf{k}} \, +\,  U \sum_\mathbf{r} \, n_{\uparrow \mathbf{r}}\,
n_{\downarrow \mathbf{r}} \; .
\end{equation}
Here $a^{\dagger}_{\sigma \mathbf{k}}$ is the fermion creation operator
with the quasi-momentum ${\bf k}$ lying in the first Brillouin zone,
$n_{\sigma \mathbf{r}} = a^{\dagger}_{\sigma \mathbf{r}}a^{\phantom{\dagger}}_{\sigma \mathbf{r}}$,
$\sigma = \uparrow \; , \downarrow$ is the spin projection,
and $\mathbf{r}$ is the integer radius vector on the d-dimensional lattice.
The lattice dispersion is given by $\varepsilon_{\mathbf{k}} = -2t \sum_{\alpha=1}^d {\rm cos}({\rm k}_{\alpha}a)$,
with $a$ and $t$ being the lattice spacing and the hopping amplitude, respectively;
$\mu_\sigma$ is the chemical potential for the component $\sigma$,
and $U$ is the on-site interaction strength. Units are chosen such that $a$ and $t$ are equal to unity.

%%%%%%%%%%%%%%%%%%%%%%%%%%%%%%%%%%%%%%%%%%%%%%%%%%%%%%%%%%%%%%%%%%%%%%%
\begin{figure}[tbh]
\sidecaption
\includegraphics[width=0.3\columnwidth,keepaspectratio=true]{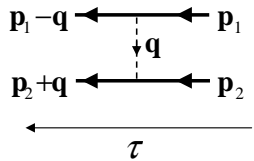}
\caption{
The diagram structural element. The dashed vertical line
represents the interaction potential. The upper (lower) solid lines
represent spin-up (spin-down) fermionic propagators.}
\label{f1}
\end{figure}
%%%%%%%%%%%%%%%%%%%%%%%%%%%%%%%%%%%%%%%%%%%%%%%%%%%%%%%%%%%%%%%%%%%%%%%%

We utilize the standard Matsubara diagrammatic technique
\cite{Fetter_Walecka}. The diagrams consist of (vertical) dashed
lines standing for the pair interaction potential $U$ and solid lines
representing particle propagators $G^{(0)}$ (see Fig.~\ref{f1}). We adopt a
convention that the imaginary-time axis is horizontal, and is
directed from right to left. Since in our case there are no
interactions within one and the same component, we require that
the upper (lower) end of each dashed line corresponds to the
spin-up (spin-down) component. With this rule, the spin-up lines
are distinguished from spin-down ones without explicit labeling.

A free fermionic propagator $G^{(0)}_{\sigma} (\mathbf{k},\, \tau=\tau_2-\tau_1)$
associated with each particle line running from $\tau_1$ to $\tau_2$ is defined by
\begin{equation}
 G^{(0)}_{\sigma} (\mathbf{k}, \tau)\,  =\,  \left\{
   \begin{array}{ll} -
     e^{-(\varepsilon_{\mathbf{k}}-\mu_{\sigma})\tau} (1-n^{(0)}_{\sigma \mathbf{k}})\, , & \textrm{~~if $\tau > 0$}\; ,\\
 + e^{-(\varepsilon_{\mathbf{k}}-\mu_{\sigma})\tau} n^{(0)}_{\sigma \mathbf{k}}\, , & \textrm{~ if $\tau < 0$}\; ,\\ \end{array} \right.
\end{equation}
with $n^{(0)}_{\sigma \mathbf{k}} = \left[ 1+e^{\beta
(\varepsilon_{\mathbf{k}}-\mu_{\sigma})} \right]^{-1}$ being the
occupation of the state $(\sigma, \mathbf{k})$ at inverse
temperature $\beta$ for free fermions on the lattice. A Green's
function with equal time variables (a closed fermion loop) is
understood as $G^{(0)}_{\sigma} (\mathbf{k}, \, \tau = -0)$.
To obtain the right weight and sign, one also has to ascribe the factor
$(-1)^{N+N_l}/(2\pi)^{Nd}$ to each diagram of order $N$, with $N_l$ being
the number of closed fermion loops and $d$ the dimensionality of
the problem.

All the physical information we will need is contained in the
self-energy $\Sigma_\sigma$ \cite{Fetter_Walecka}. In analytic
treatments, it is convenient to have $\Sigma_\sigma$ in the
momentum--imaginary-frequency representation, so that the Green's
function is obtained from the Dyson equation by simple algebra:
\begin{equation}
\left[G_\sigma( \mathbf{p}, \xi)\right]^{-1}\,  =\,
\left[G^{(0)}_\sigma( \mathbf{p}, \xi) \right]^{-1} -
\Sigma_\sigma( \mathbf{p}, \xi)\; . \label{G}
\end{equation}
Numerically, we find it more appropriate to work in the
momentum--imaginary-time representation, to avoid dealing with
poles. This does not create a problem with finding $G_\sigma(
\mathbf{p}, \xi)$, since $\Sigma_\sigma( \mathbf{p}, \xi)$ is
readily obtained from $\Sigma_\sigma( \mathbf{p}, \tau)$ by a (fast)
Fourier transform.

The class of diagrams contributing to $\Sigma_\sigma( \mathbf{p},
\tau)$ is defined by the following requirements: (i) Each diagram
has two special vertices having one open fermionic end.
These vertices are separated by the time interval $\tau$ and their
open ends have the same spin projection $\sigma$ and momentum
$\mathbf{p}$ which enters the diagram at one open end and exits
at the other. (ii) Each diagram is connected. (iii) Each diagram
is irreducible, i.e., it can not be split into two disconnected
parts by cutting only a single fermionic line.

\section{Updates}

For the sake of algorithmic elegance, we choose to work with
closed-loop diagrams that (formally) have no free ends. To collect
statistics for $\Sigma$ in this representation, we fix one
propagator, say going from vertex 1 to vertex 2, and set its value
to $1$.
Below, we refer to this special propagator as `measuring'
propagator. Then, if the propagator's momentum is $\mathbf{p}$ and
the vertices have times $\tau_1, \tau_2$ respectively, we are
measuring $\Sigma( \mathbf{p}, \tau_1-\tau_2)$.
Note that $\Sigma( \mathbf{p}, \tau)$ is an anti-periodic function of $\tau$ (with $\tau=\tau_1-\tau_2$),
so we can restrict ourselves to collecting statistics for $0 < \tau < \beta$.
For a given
diagram, changing the measuring propagator---without changing the
diagram structure---is straightforwardly done by the \textit{Swap}
update.

\textit{Swap.} The update swaps the measuring propagator with one
of the regular propagators. The new measuring
propagator is chosen at random, and its value is set to $1$. Correspondingly, the
physical value of the old measuring propagator is restored, so
that it becomes just a regular propagator.

Assume that the initial measuring propagator has to be replaced with
$G^{(0)}_\alpha(\mathbf{p}, \tau)$,
after we swap it with the propagator $G^{(0)}_\beta(\mathbf{p}', \tau')$.
The acceptance ratio is given by
\begin{equation}
  P_{{\rm swap}}\,  =\,  |G^{(0)}_\alpha(\mathbf{p}, \tau)/G^{(0)}_\beta(\mathbf{p}',\tau')|\; .
\end{equation}

The problem of generating diagrams for $\Sigma$ consists of two
main tasks:  one should be able to (i) change the structure of the
diagrams, and (ii) make sure that the momentum conservation is
satisfied in every vertex. We develop a scheme that fulfills these
tasks by performing only \textit{local} updates of the diagram.
The idea---in the spirit of the worm algorithm---is to introduce
non-physical diagrams in which momentum conservation is violated
in some special vertices (called worms). The smallest required
number of such vertices is two. All updates changing the topology
of the diagram involve worms. When worms are deleted from the diagram
we return to the physical subspace.

\begin{figure}[tbh]
\includegraphics[width=0.97\columnwidth,keepaspectratio=true]{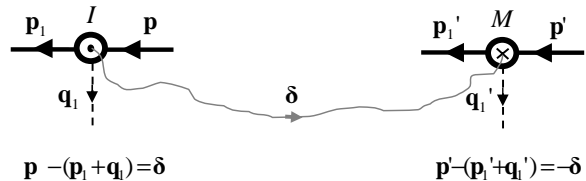}
\caption{Worms. Graphically, we can picture the worms as vertices
with \textit{conserving} momentum, by connecting the two with an
extra line (thread) that transports the momentum $\vec{\delta}$
from $\mathcal{I}$ to $\mathcal{M}$. The indistinguishability of
$\mathcal{I}$ and $\mathcal{M}$ is seen from the fact that
swapping the labels $\mathcal{I}$ and $\mathcal{M}$, and
replacing $\vec{\delta}\to -\vec{\delta}$ yields the same original
diagram.}
\label{f2}
\end{figure}

To be specific, we introduce the following notation. One of the
three-point vertices in which momentum conservation is violated,
if any, is labeled $\mathcal{I}$.
There is an excess momentum $\vec{\delta}$ associated with
$\mathcal{I}$, which is defined as the difference between the
incoming and the outgoing momenta in this vertex (see
Fig.~\ref{f2}). If present, $\mathcal{I}$ always has one, and only one,
conjugate vertex $\mathcal{M}$ with an excess momentum $-\vec{\delta}$,
as shown in Fig.~\ref{f2}. Clearly, the
distinction between $\mathcal{I}$ and $\mathcal{M}$ is merely
conventional, since replacing $\vec{\delta}$ with $-\vec{\delta}$
interchanges $\mathcal{I}$ and $\mathcal{M}$.
There is an important property of the two worms: once some path
from $\mathcal{I}$ to $\mathcal{M}$ is known, one can simply
remove the worms by propagating the excess momentum
$\vec{\delta}$ along the path.

We proceed now with the description of the updates.  The
\textit{Create/Delete} pair switches between the physical and
non-physical (worm) diagrammatic spaces by creating/deleting
$\mathcal{I}$, $\mathcal{M}$. Its role is also
to update the diagram's momenta. The \textit{Add/Remove} pair changes
the diagram order. In these updates, we add (remove) a vertex and
delete (create) $\mathcal{I}$, $\mathcal{M}$  on different spin
components at the same time. The self-complementary updates \textit{Move} and
\textit{Reconnect} are responsible for moving the worms in the
diagram and changing its topology, respectively.

\begin{figure}[tbh]
\includegraphics[width=0.97\columnwidth,keepaspectratio=true]{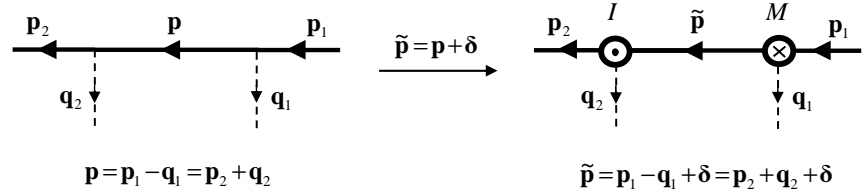}
\caption{Create update. }
\label{f3}
\end{figure}

\textit{Create.} This update is possible only---being rejected
otherwise---when we are in the physical sector ($\Sigma$-sector),
that is when $\mathcal{I,M}$ are absent. We introduce $\mathcal{I,M}$
by selecting a propagator (line) at random and adding a momentum
$\vec{\delta}$ to it (see Fig.~\ref{f3}).
[Similarly, $\mathcal{I}$ and $\mathcal{M}$ can be also created on the two vertices of an interaction line.]

Creating two worms takes the diagram from the $\Sigma$-sector to
the non-physical $W$-sector. We define the $W$-diagram weight by exactly
the same rules as for physical ones up to an arbitrary numeric factor,
so that the acceptance ratio for \textit{Create} is
 \begin{equation}
   P_{\rm create} \, =\,
 2N\, C_N\,   G^{(0)}(\mathbf{p}+\vec{\delta},\tau)/(2\pi )^d W(\vec{\delta})~G^{(0)}(\mathbf{p},\tau)\;  , \label{qcreate}
 \end{equation}
where $2N$ is the total number of propagators (including the measuring
propagator) and $W(\vec{\delta})$ is the normalized distribution from which
$\vec{\delta}$ is drawn. The variable $C_N$ is an extra weighing factor
which is assigned to a diagram of order $N$ whenever the worm is present.
To make transitions between the $\Sigma$- and $W$-sectors efficiently,
we choose $\, C_N \, =\,  C  / 2 N$, where $C$ is a constant.

The simplest choice for $W(\vec{\delta})$ is to assume a uniform
distribution in the first Brillouin zone, $W(\vec{\delta}) =
1/(2\pi)^d$. Since $\mathcal{I}$ and $\mathcal{M}$ are indistinguishable,
we find it convenient to require that the leftmost worm
is labeled $\mathcal{I}$, and the rightmost one $\mathcal{M}$
(see Fig. \ref{f3}).

The rest of the updates (except for \textit{Remove}) apply only---being
automatically rejected otherwise---when $\mathcal{I,M}$ are
present.

\textit{Delete.} Here, we first check if $\mathcal{I}$ and
$\mathcal{M}$ are connected by a single line and proceed only if they are.
We propose to remove $\mathcal{I}$ and $\mathcal{M}$ by adding
$-\vec{\delta}$ ($+\vec{\delta}$) to the momentum of the connecting line
when it is incoming (outgoing) for $\mathcal{I}$.
If the line connecting $\mathcal{I}$ and $\mathcal{M}$ is a
propagator, the acceptance
ratio for \textit{Delete} is given by the inverse of
Eq.~(\ref{qcreate}),
\begin{equation}
   P_{\rm delete}\, =\,(2\pi)^d W(\vec{\delta})\, G^{(0)}(\mathbf{p},\tau)
  \,  /\, 2N\, C_N\, G^{(0)}(\mathbf{p} \pm \vec{\delta},\tau) \; .
\label{qdelete}
\end{equation}

Care should be taken in the special case when
$\mathcal{I}$ and $\mathcal{M}$ are connected by two lines
(forming a closed fermion loop). A factor  of $2$ $[1/2]$ should
be included in Eq.~(\ref{qdelete}) [Eq.~(\ref{qcreate})], because
the excess momentum $\vec{\delta}$ can be attributed
to any of the two lines with probability $1/2$.

\begin{figure}[tbh]
\includegraphics[width=0.97\columnwidth,keepaspectratio=true]{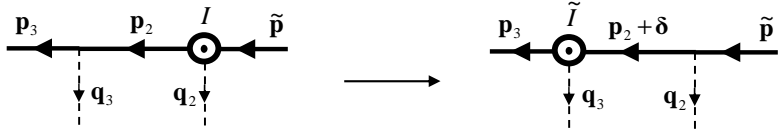}
\caption{Example of Move left update.}
\label{f4}
\end{figure}

\textit{Move.} We can move both $\mathcal{I}$ and $\mathcal{M}$
along the lines through the whole diagram. Moving a worm along a
line is only allowed if there is no other worm on this line.
[Otherwise, the update is equivalent to \textit{Delete.}] For
example, to move $\mathcal{I}$ left (Fig.~\ref{f4}) we must add
$\vec{\delta}$ to the line on the left from $\mathcal{I}$. This
will restore the momentum conservation in $\mathcal{I}$ and create
a momentum discrepancy $\vec{\delta}$ in the vertex left from
$\mathcal{I}$. This is our new $\mathcal{I}$, temporarily denoted
as $\tilde{\mathcal{I}}$, to avoid confusion with the old one.

The acceptance ratio of this move is given by
\begin{equation}
  P_{\rm move}\, =\, G^{(0)}(\mathbf{p}_2 +
\vec{\delta}, \tau)/G^{(0)}(\mathbf{p}_2, \tau)\; ,
\end{equation}
with $G^{(0)}(\mathbf{p}_2, \tau)$ the outgoing propagator of
$\mathcal{I}$ before the update. Moving $\mathcal{I}$ to the left
or right is chosen with equal probability.

Graphically, we can picture the worms as vertices with
conserving momentum, provided we draw an artificial thread directly
connecting the worms and transporting the momentum $\vec{\delta}$
from $\mathcal{I}$ to $\mathcal{M}$ (see Fig.~\ref{f2}). In this
language, removing the worms means merging the thread with some
path formed by the lines connecting the worms and adding the thread momentum
to their momenta.
If only a part of the connecting path is merged with the thread, this
results in moving the worms. This way of thinking visually tells
us whether we have to add or subtract $\vec{\delta}$ on lines
while moving the worms.

\begin{figure}[tbh]
\includegraphics[width=0.97\columnwidth,keepaspectratio=true]{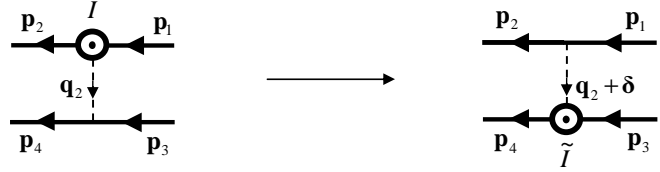}
\caption{Example of Move down update.}
\label{f5}
\end{figure}

Moving $\mathcal{I},\mathcal{M}$ ``up'' or ``down'' along an
interaction line, we switch to a different spin component. For
example, to move $\mathcal{I}$ down (up) (see Fig.~\ref{f5}), we
have to add $\vec{\delta}$ ($-\vec{\delta}$) to the dashed line
(interaction) momentum. The acceptance ratio for moving
$\mathcal{I}$ down is given by
\begin{equation}
P_{\rm move}\,  =\, U(\mathbf{q}_2+\vec{\delta})/U(\mathbf{q}_2)\;
.
\end{equation}

Obviously, the move right/left/up/down updates are the same for
$\mathcal{M}$ with the only difference that instead of adding $\pm
\vec{\delta}$ to the lines we have to subtract it.

\textit{Reconnect.} This simple update can be
performed---and is automatically rejected otherwise---only if
$\mathcal{I}$ and $\mathcal{M}$ occupy the \textit{same} spin
component.  One swaps the incoming end of $\mathcal{I}$ with that
of $\mathcal{M}$. It is important, however, that the update
changes the excess momentum $\vec{\delta}$ associated with the
worms. [It is also worth noting that without invoking the worms
such an update would be generally impossible by momentum conservation.]
If before the update the momenta on the incoming ends
of $\mathcal{I}$ and $\mathcal{M}$ were $\mathbf{p}_1$ and
$\mathbf{p}_2$ respectively, then after the update the new excess
momentum is given by $\tilde{\vec{\delta}}=\vec{\delta}
+\mathbf{p}_2-\mathbf{p}_1$.

 The acceptance ratio for \textit{Reconnect} is
 \begin{equation}
   P_{\rm rec} \, =\,  \left|{G^{(0)}(\mathbf{p}_1, \tau_{\mathcal{M}}-\tau_1)~G^{(0)}(\mathbf{p}_2, \tau_{\mathcal{I}}-\tau_2)
\over G^{(0)}(\mathbf{p}_1,
\tau_{\mathcal{I}}- \tau_1)~G^{(0)}(\mathbf{p}_2, \tau_{\mathcal{M}}- \tau_2)}
\right|\; ,
 \end{equation}
 with $\tau_{\mathcal{I}}$, $\tau_{\mathcal{M}}$ the imaginary times of the
 vertices at which $\mathcal{I}$ and $\mathcal{M}$ are located, respectively. The
 variable $\tau_1$ ($\tau_2$) gives the imaginary time of the vertex from which
 a propagator runs to the vertex of $\mathcal{I}$ ($\mathcal{M}$) with
 momentum $\mathbf{p}_1$ ($\mathbf{p}_2$), before the update.

\begin{figure}[tbh]
\includegraphics[width=0.80\columnwidth,keepaspectratio=true]{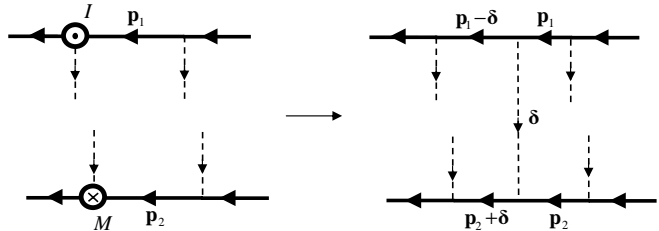}
\caption{Add update.}
\label{f6}
\end{figure}

\textit{Add.} This update, adding a new interaction line, is only
possible---being automatically rejected otherwise---if
$\mathcal{I}$ and $\mathcal{M}$ have
\textit{different} spin components. For definiteness, we insert
the new interaction line between the propagator coming into
$\mathcal{I}$ and the one coming into $\mathcal{M}$, as in
Fig.~\ref{f6}. At a randomly chosen time, these lines are
broken and the four-point vertex is inserted in the breaks. In
this particular realization, there is no freedom in choosing the
momentum along the new dashed line: we set it equal to $\vec{\delta}$,
which means that the update inevitably deletes $\mathcal{I}$ and
$\mathcal{M}$.

The acceptance ratio for \textit{Add} is
 \begin{eqnarray}
   P_{\rm add} & = & \frac{U(\vec{\delta})}{(N+1)~W_1(\tau)~C_N}
\left| \frac{G^{(0)}_{\uparrow}(\mathbf{p}_1-\vec{\delta},\tau_{\mathcal{I}} - \tau)~G^{(0)}_{\uparrow}(\mathbf{p}_1,\tau-\tau_1)}
 {G^{(0)}_{\uparrow}(\mathbf{p}_1, \tau_{\mathcal{I}} - \tau_1)} \right|
 \nonumber \\
&  &  \times \left| \frac{G^{(0)}_{\downarrow}(\mathbf{p}_2+\vec{\delta},
\tau_{\mathcal{M}} - \tau)~G^{(0)}_{\downarrow}(\mathbf{p}_2,
\tau - \tau_2)}{G^{(0)}_{\downarrow}(\mathbf{p}_2,
\tau_{\mathcal{M}}-\tau_2)} \right| \; , \label{qadd}
 \end{eqnarray}
where $N$ is the order of the diagram and $W_1(\tau)$ is the distribution
from which the time $\tau$ of the new interaction vertex is drawn.
 We have assumed that $\mathcal{I}$ occupies the spin-up component, without
 loss of generality.

\textit{Remove.} This update is a straightforward inverse of the
previous one. It can be performed (accepted) only in the
$\Sigma$-sector. It simultaneously removes an interaction line and
creates a pair of worms. If the removed line had momentum
$\mathbf{q}$, for $\mathcal{I}$ and $\mathcal{M}$ we shall have
$\vec{\delta}=\mathbf{q}$.

In view of the indistinguishability of $\mathcal{I}$ and
$\mathcal{M}$, we require that $\mathcal{I}$ ($\mathcal{M}$) is
created on the spin-up (spin-down) line. The acceptance ratio of
this move is given by the inverse of Eq. (\ref{qadd}) with $N$
being the final diagram order now. Adding or
removing an interaction line can also involve a change of the
measuring propagator when this propagator is the incoming propagator
of $\mathcal{I}$ or $\mathcal{M}$.

\section{Useful Tricks and Relations} \label{sec:tricks}

\textbf{Connectivity and irreducibility. } The updates discussed
above are constructed in such a way that disconnected physical
diagrams are never sampled. In the $\Sigma$-sector, the only
update that can change the topology of the diagram is the
\textit{Swap} move. However, swapping the measuring propagator
cannot create a disconnected piece. When we are dealing with a
diagram in the $W$-sector, \textit{Add-Remove} and
\textit{Reconnect} are the only updates that can change the
topology of the diagram. These updates can create two disconnected
pieces in the $W$-sector. In this case, however, each disconnected
piece will contain one worm end. Going back to the $\Sigma$-sector
amounts to bringing the worm ends together via \textit{Add} or
\textit{Reconnect}, at the same time restoring the connectivity.

To keep track of the irreducibility of the diagrams, we use a
hash table of lines momenta. The idea is
that whenever the diagram is reducible, there is at least one
propagator which carries the total momentum of the diagram. The
reducibility of the diagram can then be established easily
by checking the hash table and finding a momentum equal to that
of the measuring propagator. We also check irreducibility right before
each measurement of the self-energy.

\textbf{Bold propagators.} The purpose of this trick is to reduce
the space of diagrams sampled by Monte Carlo. As already
mentioned, this becomes essential when one is dealing with a
sign-alternating series, in which the sign problem scales
exponentially with the number of diagrams. Basically, the idea
follows from Dyson's equation, which allows one to reconstruct the
complete function $G$ from its elementary building blocks
$\Sigma$. Next, in the series for $\Sigma$, one has to eliminate
all diagrams already accounted for by the replacement $G^{(0)} \to
G$ done for all propagator lines. At this point the scheme becomes
self-consistent. In addition, one can employ geometrical series to
define screened interaction lines, and, correspondingly, eliminate
all diagrams with simple fermionic loops. At the time of writing
we are in the process of implementing the full version of the
bold-line trick.

In the simplest implementation, we build the diagrams on a modified Green's
function $\tilde{G}$ instead of $G^{(0)}$. The function $\tilde{G}$
is obtained from $G^{(0)}$ by incorporating the lowest order ``tadpole'' diagram
(i.e., the mean-field solution) self-consistently.
This reduces the number of sampled diagrams, since all diagrams that can be
split into two disconnected parts by cutting a single interaction line should
be left out, being already taken into account in the new Green's function.
Mathematically, $\tilde{G}$ is obtained from Eq.~(\ref{G}) by simply
shifting the chemical potential, $\mu_{\sigma} \longrightarrow
\mu_{\sigma}-\Sigma^{(1)}_\sigma$, where $\Sigma^{(1)}_\sigma = U
n_\sigma$ is the self-energy in the lowest order and $n_\sigma$ is the
density of the component $\sigma$. In principle, updates
creating tadpole-contributions can simply be rejected. However, to
keep the Markov sampling of diagrams ergodic, the presence of a
limited number of closed propagators $\tilde{G}_{\sigma}(\mathbf{k},
\tau=-0)$ is allowed, but such diagrams are excluded from the statistics of $\Sigma_\sigma$.

\textbf{Resummation of the diagrams. } The use of resummation
techniques is a crucial ingredient of the diagrammatic Monte Carlo
approach \cite{fermipolaron}. We found that the resummation methods
effectively reduce the error bars, and improve
convergence of the Monte Carlo results. In general, for any
quantity of interest---in our case, self-energy---one constructs
partial sums
\begin{equation}
\Sigma (N_*)\, =\,  \sum_{N=1}^{N_*} D_N F_N^{(N_*)} \; ,
\label{partials}
\end{equation}
defined as sums of all terms up to order $N_*$ with the $N$-th
order terms being multiplied by the factor $F_N^{(N_*)}$, which has a step-like form as a function of N: in the
limit of large $N_*$ and $N \ll N_* $ the multiplication factors
$F$ approach unity while for $N \to N_* $ they suppress
higher-order contributions in such a way that
the series $\sum_{N=1}^{\infty} D_N F_N^{(N_*)}$ becomes convergent.
The only other requirement is that the crossover region from unity
to zero has to increase with $N_*$. There are infinitely many ways
to construct multiplication factors satisfying these conditions.
And this yields an important consistency check: final results
have to be independent of the choice of $F$. In the Ces\`aro-Riesz
summation method we have
\begin{equation}
F_N^{(N_*)}\, =\, [(N_*-N+1)/N_*]^\delta \;, ~~~~~
\mbox{(Ces\`aro-Riesz)}\; . \label{factor1}
\end{equation}
Here $\delta > 0$ is an arbitrary parameter ($\delta = 1$
corresponds to the Ces\`aro method). The freedom of choosing the
value of Riesz's exponent $\delta$ can be used to optimize the
convergence properties of $\Sigma (N_*)$. Empirically it was found
that the factor
\begin{equation}
F_N^{(N_*)} \, =\, C^{(N_*)}\sum_{m=N}^{N_*}\exp{\left[-(N_*+1)^2
/ m(N_*-m+1) \right] } \; , \label{factor2}
\end{equation}
where $C^{(N_*)}$ is such that $F_1^{(N_*)}=1$, often leads to a faster
convergence \cite{fermipolaron}.

We proceed as follows. With the series truncated at order $N_*$,
we determine the physical quantity of interest (say, number
density or energy), and then extrapolate its dependence on $N_*$
to the $N_* \to \infty$ limit.

\textbf{Density and energy.} The density is obtained from the
Green's function through the standard relation
\begin{equation}
  n_{\sigma} \, =\, \int_{BZ} {d\mathbf{k}\over (2\pi)^d}\,  G_{\sigma}(\mathbf{k},
  \tau=-0)\; ,
\end{equation}
where the integration is over the first Brillouin zone.
Analogously, the kinetic (hopping) energy is found as
\begin{equation}
E^{(\mathrm{kin})}_{\sigma}/V \, =\,
 \int_{BZ}  {d\mathbf{k}\over (2\pi)^d} \,
\varepsilon_{\mathbf{k}} G_{\sigma}(\mathbf{k},   \tau=-0)\; ,
\end{equation}
and the potential (interaction) energy is obtained via
\begin{equation}
2E^{(\mathrm{pot})}_{\sigma }/V\, =\,   {\rm lim}_{\tau' \to \tau+0} \int_{BZ}
{d\mathbf{k}\over (2\pi)^d} \bigg[ -\frac{\partial}{\partial\tau}
-
  \varepsilon_{\mathbf{k}} + \mu_{\sigma}
  \bigg]G_{\sigma}(\mathbf{k},\tau-\tau')\; ,
\end{equation}
where $V$ is the volume of the system.

\section{Illustrative Results} \label{sec:results}

To illustrate the application of DiagMC, we simulated the
equations of state---density and energy as functions of chemical
potential and temperature---in one (1D) and three (3D) dimensions.
Diagrammatically, there is no qualitative difference between 1D
and higher-dimensional cases. Meanwhile, in 1D, where fermions are
equivalent to hard-core bosons, we have the advantage of comparing
DiagMC results with very accurate answers obtained with the bosonic
worm algorithm \cite{Barbar_Gunes}.

By the (discussed in the Introduction) Dyson argument, we expect
that at any finite temperature the thermodynamic functions are
analytic functions of $U$ in a certain vicinity of the point
$U=0$, because this point is not singular in the case of the {\it
discrete} fermionic system. Correspondingly, the expansion in
powers of $U$ (that, by dimensional analysis, can be understood as
the expansion in powers of the dimensionless parameter $U/T$) is
supposed to be convergent at small enough $U$, or, equivalently,
large enough $T$. In 1D there are no phase transitions at any
finite $T$, so that one can expect that the expansion in powers of
$U$ is convergent (or at least resummable) down to arbitrarily
low temperatures. In higher dimensions, an essential
non-analyticity of thermodynamic functions appears due to phase
transitions. Hence, the simplest version of DiagMC described in
this work is expected to work only at high enough temperatures.

Note that employing resummation techniques is important not only
for extending the method to divergent series, or improving the series
convergence properties, but also for estimating systematic errors
due to extrapolation from a finite number of terms we are able to calculate.

\begin{figure}[tbh]
\includegraphics[width=0.5\columnwidth,keepaspectratio=true]{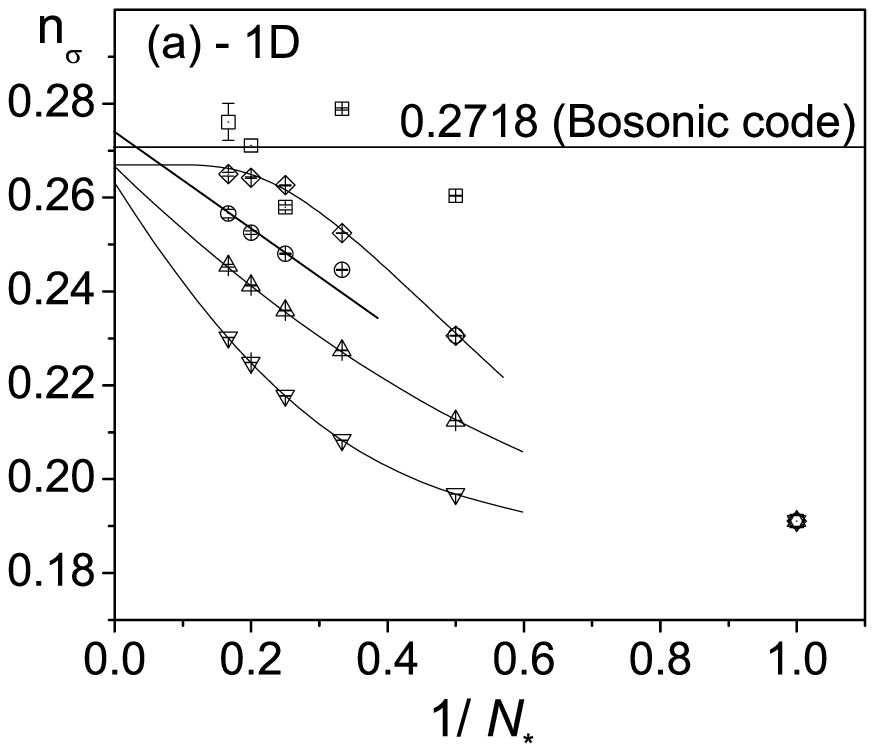}
\includegraphics[width=0.5\columnwidth,keepaspectratio=true]{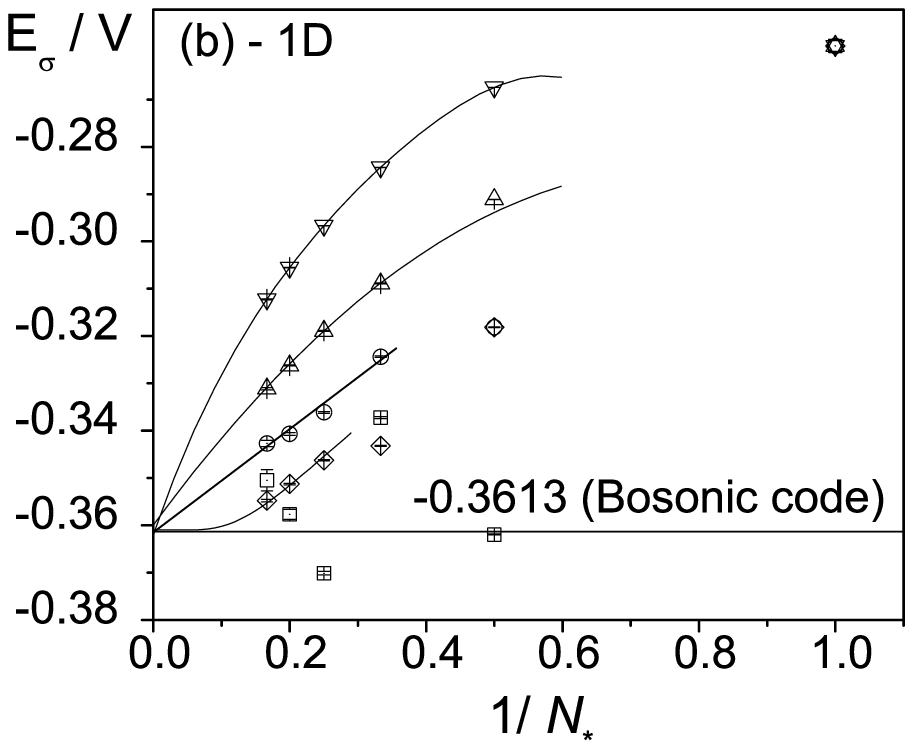}
\\
\includegraphics[width=0.5\columnwidth,keepaspectratio=true]{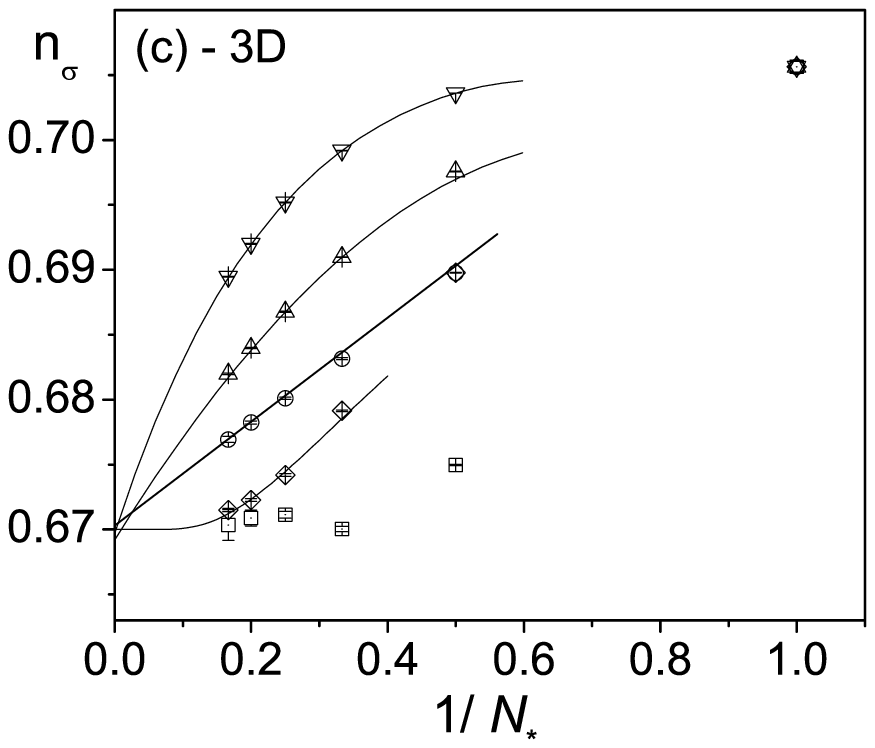}
\includegraphics[width=0.5\columnwidth,keepaspectratio=true]{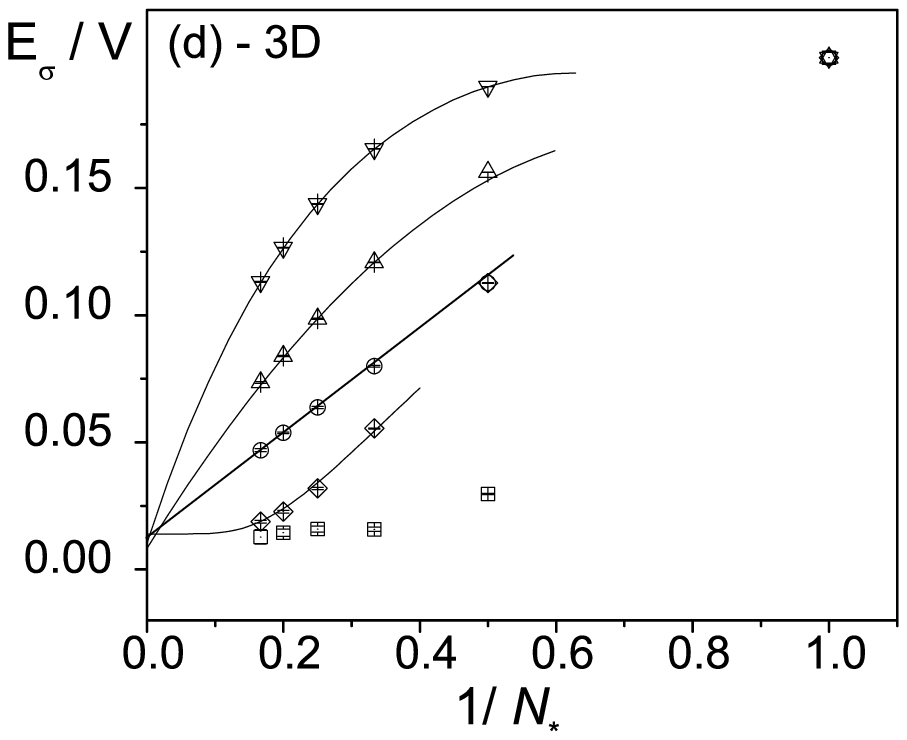}
\caption{The number density $n_{\sigma}$ and the total energy
density $E_{\sigma}/V$ per spin component in 1D [panels (a), (b)]
with $U=4.0$, $\mu_{\uparrow, \downarrow}=-0.5$, and $T=0.3$, and
in 3D [panels (c), (d)] with $U=4.0$, $\mu_{\uparrow,
\downarrow}=1.5$, and $T=0.5$ plotted as  functions of the inverse
of the maximum diagram order $N_*$. The exact results in 1D,
$n_{\sigma}=0.2718$ and $E_{\sigma}/V=-0.3613$, were obtained from
simulations of two-component bosons \cite{Barbar_Gunes}. The
results of different resummation procedures are shown---Ces\`{a}ro
sum (circles), Ces\`{a}ro-Riesz with $\delta=2$ (triangles),
Ces\`{a}ro-Riesz with $\delta=4$ (inverted triangles), and
resummation with $F^{N_*}_{N}$ given by Eq.~(\ref{factor2})
(diamonds), and bare series without resummation (squares). The
solid lines show the best fits [linear for Ces\`{a}ro, polynomial
of order $\delta$ for Ces\`{a}ro-Riesz, and exponential for
Eq.~(\ref{factor2})].}
\label{fig:1D}
\end{figure}

The simulation results (plotted versus the inverse of the maximum
diagram order $N_*$) for 1D and 3D and different resummation
methods are shown in Fig. \ref{fig:1D}. In 1D the horizontal
straight lines in both energy and density plots mark the exact
answers obtained from simulations of two-component bosons
\cite{Barbar_Gunes}. The computational effort in 3D was about
$4800$ CPU-hours. Note that the error bars in the
final answer are dominated by the systematic $N_* \to \infty$
extrapolation errors which are estimated from the spread of
results for different extrapolation fits and resummation methods.
In particular, for density in 1D the relative uncertainty due to
extrapolation is of the order of $5 \%$, suggesting that going to
higher $N_*$ is desirable, which is only possible with further
implementation of the bold-line tricks.

\section{Conclusion}

We have developed a diagrammatic Monte Carlo scheme for a system
of interacting fermions. For illustrative results, we have
obtained equations of state (density and energy as functions of
chemical potential and temperature) for the repulsive Hubbard
model. In its current version, the scheme applies to the range of
temperatures above the critical point. Generalizations of the
scheme to temperatures below the critical point are possible by
using one of the following two strategies (as well as their
combination). (i) A small term explicitly breaking the symmetry
(relevant to the critical point) can be introduced into the
Hamiltonian. The phase transition then becomes a crossover, and
the convergence/resummability of series is likely to take place at
any temperature. (ii) Anomalous propagators can be introduced. The
latter trick naturally involves the bold-line technique which is
important on its own, even above the critical point, since it
reduces the number of diagrams and thus alleviates the sign
problem.

Tolerance to the sign problem is the single most important feature
of DiagMC. One can formulate the sign problem as the impossibility of obtaining
results with small error bars for system sizes which allow a reliable
and controlled extrapolation to the thermodynamic limit. Within the
DiagMC approach the thermodynamic limit is obtained for free, while, as
demonstrated in this work, an extrapolation to the infinite diagram
order can be done sensibly before the error bars explode.

\section{Acknowledgements}

The work was supported by the National Science Foundation under
Grant PHY-0653183, the Fund for Scientific Research - Flanders
(FWO), and by the DARPA OLE program.
Part of the simulations ran on the Hreidar 
cluster at ETH Z\"urich.

%INDEX%%%%%%%%%%%%%%%%%%%%%%%%%%%%%%%%%%%%%%%%%%%%%%%%%%%%%%%%%%%%%%%
% Please check with the editor of your book whether he plans to
% include a "mutual" subject index - if so, please code your entries
% in the standard syntax. For your own purposes you may print your
% "personal" index by using the following commands:
%
%\clearpage
%\addcontentsline{toc}{section}{Index}
%\flushbottom
%\printindex
%%%%%%%%%%%%%%%%%%%%%%%%%%%%%%%%%%%%%%%%%%%%%%%%%%%%%%%%%%%%%%%%%%%%%


\begin{thebibliography}{8.}
\bibitem{Barbar_Gunes} The worm code for the two-component Bose
Hubbard model is the courtesy of Barbara Capogrosso-Sansone and
\c{S}ebnem G\"{u}ne\c{s} S\"{o}yler.
\end{thebibliography}
\end{document}